\documentstyle[twocolumn,aps,epsfig]{revtex}
\begin{document}
\draft
\title{Gravitational Decay Modes of the Standard Model Higgs Particle}
\author{Y.N. Srivastava }
\address{Physics Department \& INFN, University of Perugia, Perugia, Italy}
\author{A. Widom}
\address{Physics Department, Northeastern University, Boston MA U.S.A.}
\maketitle
\begin{abstract}
If the Einstein field equations are employed at the tree 
level, then the decay of the standard model Higgs particle 
into two gravitons is shown to be independent of the 
gravitational coupling strength G. The result follows from 
the physical equivalence between the Higgs induced 
``inertial mass'' and the ``gravitational mass'' of general 
relativity. If the Higgs mass lies well between the mass of 
a bottom quark anti-quark pair and the mass of a top quark 
anti-quark pair, then the Higgs decay into two gravitons 
will dominate both the QED induced two photon decay and the 
QCD induced two jet decays.
\end{abstract}  
\narrowtext

The last major notion of the standard electro-weak model\cite{1,2,3} which 
has yet to receive experimental confirmation\cite{4,5,6,7,8,9} 
is the prediction of the Higgs particle\cite{10,11,12,13,14}. 
In part, the problem may be simply connected to 
our lack of knowledge of the value of the Higgs mass  
\begin{math} M_H \end{math}\cite{15}. But other problems arise 
on a more conceptual level. 

The Higgs field is thought to provide the mechanism for the existence 
of all {\em inertial} mass\cite{16}. Yet the standard model does not 
relate this Higgs induced inertial mass to the important and 
presumably equivalent value of the  {\em gravitational} mass. 
In what follows, we shall add to the standard electro-weak model Higgs 
notion of inertial mass\cite{17}, the Einstein notion of gravitational 
mass via the conventional curvature field equations of general 
relativity 
\begin{equation}
R_{\mu \nu}-\left({1\over 2}\right)g_{\mu \nu}R=
\left({8\pi G\over c^4}\right)T_{\mu \nu}.
\end{equation}
In particular, for the trace 
\begin{math} T=g^{\mu \nu }T_{\mu \nu}  \end{math}
we shall employ  
\begin{equation}
T=-\left({c^4\over 8\pi G}\right)R .
\end{equation}

In the standard model, the Higgs field is entirely responsible 
for the possible existence of \begin{math} T\ne 0  \end{math}.  
Thus, the Higgs field is entirely responsible for the 
possible existence of a non-trivial scalar curvature 
\begin{math} R\ne 0  \end{math} in general relativity. 
Nevertheless, the modes of the interaction between the Higgs field 
and conventional Einstein gravity have gone virtually unnoticed with 
regard to high energy laboratory quantum gravity experiments.

The reason for this sad state of affairs is that the value of 
the Planck mass \begin{math} M_P  \end{math}, defined by  
\begin{equation}
\left({GM_P^2\over \hbar c}\right)=1,
\end{equation}
is thought to be much too large to allow for quantum gravity 
observations using conventional high energy beams. The weak interaction 
(Fermi coupling \begin{math} G_F \end{math}) version of Eq.(3), i.e. 
\begin{equation}
\left({\sqrt{2}\ G_F M_F ^2\over \hbar c}\right)=1,
\end{equation} 
sets the mass scale at the vacuum condensation value of the Higgs field 
\begin{equation}
M_F=\left({\hbar \left<\phi \right>\over c}\right).
\end{equation}
The value of \begin{math} M_F \end{math} is thus known to be 
\begin{equation}
M_F \approx  246 \ GeV/c^2 ,
\end{equation}
well within the present day technology of high energy beams. 
Let us return to the quantum gravity aspects of the problem.  

For Higgs particle excitations one normally writes the total field 
\begin{equation}
\phi =\left<\phi \right>+\chi ,
\end{equation}
while the effective action employed for computing the decay of the Higgs 
particle is given by 
\begin{equation}
S_{eff}=\left({1\over c\left<\phi \right>}\right)
\int \chi T d\Omega,
\end{equation}
where \begin{math} d\Omega =\sqrt{-g}d^4x  \end{math} is the 
space-time ``volume'' element. While  \begin{math} T \end{math},  
the trace of the stress tensor, has many contributions, e.g. a term 
\begin{math}(-mc^2\bar{\psi}\psi ) \end{math}
for each massive fermion species, the total sum over {\em all} the  
fields coupling into the Higgs particle is most simply described by 
the effective action in Eq.(8). 

From Eqs.(2) and (8), it follows that the effective action 
depends on the scalar curvature 
\begin{equation}
S_{eff}=-\left({c^3\over 8\pi G\left<\phi \right>}\right)
\int \chi R d\Omega .
\end{equation} 
Eqs.(8) and (9) express the fact that the Higgs couples 
equivalently into inertial and gravitational mass, but in the 
latter case we can relate the result to the Lagrangian density 
\begin{math} {\cal L}_g \end{math} of the gravitational 
field; i.e.  
\begin{equation}
S_g = \left({c^3\over 16\pi G}\right)\int Rd\Omega 
= \left({1\over c}\right)\int{\cal L}_g d\Omega . 
\end{equation}
The Higgs coupling to Lagrangian density of gravitons is then 
\begin{equation}
S_{eff}=-\left({2\over c\left<\phi \right>}\right)
\int \chi  {\cal L}_g  d\Omega .
\end{equation}
One may now be aware that in the Higgs coupling to gravitons, 
the gravitational coupling strength \begin{math} G \end{math} 
has very quietly slipped away. (The situation is reminiscent of the 
discussions\cite{18} between Bohr and Einstein on the completeness of 
the quantum mechanical view. Toward the end of these discussions, 
Bohr had to invoke general relativity to ``save'' the energy-time 
uncertainty principle. Notwithstanding the need for a finite 
gravitational coupling \begin{math} G \end{math} in the intermediate 
stages of the argument, \begin{math} G \end{math}  dropped out of the 
final results.)

A general rule is that an oscillator Hamiltonian 
\begin{math}H = (1/2)\hbar \omega 
(aa^\dagger +a^\dagger a) \end{math} 
corresponds to an oscillator Lagrangian
\begin{math} L = (1/2)\hbar \omega 
(a^\dagger a^\dagger + aa) \end{math}. For the problem at hand, 
\begin{math} {\cal L}_g \end{math} may create 
or may destroy two gravitons. The rate at which a Higgs 
at rest will decay into two gravitons requires the matrix element 
\begin{math}\left<gg\left|S_{eff}\right|H\right>\end{math}. In 
terms of quantum fields, one requires 
\begin{math} \left<0\left|\chi \right|H\right> \end{math} and 
\begin{math} \left<gg\left|{\cal L}_g \right|0\right> \end{math}.
From Eq.(5) and (11), one computes the rate   
\begin{equation}
\Gamma (H\to g+g)=\left({1\over 16\pi }\right)
\left({M_H\over M_F}\right)^2\left({M_H c^2\over \hbar}\right).
\end{equation} 
In terms of the Fermi coupling strength Eq.(4), the Higgs into two 
gravitons has the decay rate  
\begin{equation}
\Gamma (H\to g+g)=\left({\sqrt{2}\over 16\pi }\right)
\left({G_F M_H^2\over \hbar c}\right)\left({M_H c^2\over \hbar}\right),
\end{equation} 
which is the central result of this work. The gravitational coupling 
strength does not appear in the final gravitational decay rate 
due to the physical equivalence between 
the inertial (Higgs induced) mass and the gravitational mass.

Our central Eq.(13) for the Higgs decay into two gravitons  
\begin{math} H\to g+g\end{math} may in some ways be compared 
with the computation of Higgs decay into two photons  
\begin{math}  H\to \gamma +\gamma \end{math}.
A ``scalar particle'' decay into two photons begins with the 
Schwinger anomaly\cite{19} for the trace of the stress tensor    
\begin{math} T_\gamma =(2\alpha /3\pi ){\cal L}_\gamma \end{math},  
where \begin{math} \alpha \end{math} is the quantum electrodynamic 
coupling strength, and \begin{math} {\cal L}_\gamma  \end{math} 
is the free photon Lagrangian density. 
The anomalous \begin{math} T_{\gamma } \end{math} is a one loop 
process. The one loop Higgs decay rate into two photons is lower 
than the tree level Higgs decay into two gravitons by more than   
\begin{math}\sim 10^{-6}  \end{math}. (In reality, one requires 
a one loop renormalized coupling strength 
\begin{math} \alpha \end{math}, and this  
further lowers the two photon decay rate relative to the 
two graviton decay rate.) In a similar fashion, 
the Higgs decay into two gravitons is seen to be much larger than 
the decay into two gluon jets. If the Higgs mass is much 
larger than twice the bottom quark mass but still much less than twice 
the top quark mass, then the decay of the Higgs into two gravitons 
dominates too the decay of the Higgs into quark anti-quark 
jet pairs. Lastly, if the mass of the Higgs is smaller than 
the mass of two \begin{math} Z  \end{math} Bosons or the 
mass of a \begin{math} W^+ W^-  \end{math} pair, then the Higgs 
into two graviton decay rate will dominate the decay rate into 
all \begin{math} SU(2)\times U(1) \end{math} channels; the 
heavy gauge Boson pairs are ruled out on the 
basis of the above kinematics. 

Consider the analogy between Higgs decay  
\begin{math} H\to g+g \end{math} and the well known weak interaction  
decay \begin{math} Z\to \nu + \bar{\nu} \end{math}\cite{20} The latter 
(\begin{math} Z  \end{math}-decay) has been observed even 
though the neutrino anti-neutrino pair escapes direct detection other 
than by ``missing'' the four momenta. The full process is 
\begin{math} e^- + e^+ \to Z+\gamma \to \nu + \bar{\nu}+\gamma \end{math}. 
There is a burst of  {\em soft}  photon radiation indicating that 
the electron positron pair has been destroyed, and then there is 
``nothing''. Now suppose, as one specific example among many, 
the following analog event. One produces a Higgs from a proton 
anti-proton event 
\begin{math} p^+ + p^- \to H \to g + g \end{math}.
Here too would be a soft photon radiation burst from the proton 
anti-proton destruction, and then {\em nothing} but the missing 
four momenta of the two hard final state gravitons. Such a process 
would occur resonantly at a square total four momentum 
\begin{math} s = -(P^+ + P^-)^2/c^2 = M_H^2 \end{math}. 

The experimental search for the Higgs particle has been 
argued above to also be an experimental search for quantum gravity. 
Presently there is no direct experimental evidence for the Higgs 
particle nor for quantized gravitational waves (gravitons). 
It would however appear that the two are closely connected. 
Further progress on the source of inertial masses ultimately 
requires that gravity be added to the electroweak sector. 
It is hoped that the present exploration of these ideas will 
stimulate further work on these notions.

\end{document}